# Investigation of Galactic Supernova Remnants and their Environment in 26°.6 < l < 30°.6, |b| ⩽ 1°.25 Using Radio Surveys

Tian-Xian Luo[1], Ping Zhou[1,2], and Hao-Ning He[3,4]
[1] School of Astronomy and Space Science, Nanjing University, 163 Xianlin Avenue, Nanjing 210023, People's Republic of China; pingzhou@nju.edu.cn
[2] Key Laboratory of Modern Astronomy and Astrophysics, Nanjing University, Ministry of Education, People's Republic of China
[3] Key Laboratory of Dark Matter and Space Astronomy, Purple Mountain Observatory, Chinese Academy of Sciences, Nanjing 210023, People's Republic of China
[4] Astrophysical Big Bang Laboratory, RIKEN, Wako, Saitama 351-0198, Japan
Received 2024 April 12; revised 2024 May 4; accepted 2024 May 6; published 2024 June 28

## Abstract

The problem of missing Galactic supernova remnants (SNRs) refers to the issue that the currently known Galactic SNRs are significantly incomplete compared to the theoretical prediction. To expand the sample of Galactic SNRs, we use GLEAM and THOR+VGPS data across four wave bands ranging from 118 to 1420 MHz to drive a spectral index map covering the region within 26°.6 < l < 30°.6, |b| ⩽ 1°.25, where numerous SNR candidates were recently found. By using the spectral index map of the sky region and detailed analysis of the spectral indices of individual sources, we confirmed four SNR candidates, namely G26.75+0.73, G27.06+0.04, G28.36+0.21, and G28.78−0.44, as SNRs. Additionally, we discovered an expanding molecular superbubble located in this region, discussed pulsars associated with SNR candidates, and discovered a long Hα filament that spatially overlaps with the candidate G29.38+0.10. We suggest that the problem of missing Galactic SNRs not only arises from observation limitations, but also could be due to the low-density environments of some SNRs, and the different supernova explosion properties.

*Unified Astronomy Thesaurus concepts:* Supernova remnants (1667); Radio continuum emission (1340); Superbubbles (1656)

## 1. Introduction

Supernova remnants (SNRs) are formed through the interaction between the ejected material of a supernova (SN) explosion and the surrounding interstellar medium (ISM). They are important mechanical energy sources of galaxies, enriching the galaxy with heavy metals, and regulating galaxy evolution and star formation rate (Kobayashi & Nomoto 2009; Badenes et al. 2009; Stoll et al. 2013; Vink 2020). They are also proposed as a promising origin of interstellar dust (Barlow et al. 2010; Gall et al. 2011; Matsuura et al. 2011; Gomez et al. 2009; Temim & Dwek 2013; Bevan et al. 2017; Rho et al. 2018). The study of SNRs provides insights into whether SNe, on average, act as sources or destroyers of interstellar dust (Michałowski 2015; Kirchschlager et al. 2019). SNRs are important factories of cosmic rays (CRs) via the diffusive shock acceleration processes (Malkov & Drury 2001), while it remains uncertain whether SNRs are the primary contributors to Galactic CRs. To quantify the aforementioned contributions of SNRs to the galaxy, it requires a a good understanding of the SNR number and population in a galaxy.

Theoretical and statistical studies of Galactic SNRs suggest that the currently known SNRs are far from complete. According to the estimated distances of some known SNRs and some other statistical parameters, the number of Galactic SNRs is expected to be ∼1000 (Li et al. 1991). A more recent study (Ranasinghe & Leahy 2022) with improved SNR distances suggests that the number could be in the range ∼2400–5600. It is much higher than the ∼300–400 known Galactic SNRs listed in the SNR catalogs (Ferrand & Safi-Harb 2012; Green 2022). Therefore, there are still many Galactic SNRs waiting to be identified.

Many known Galactic SNRs are first identified through radio observations because most are visible in the radio band during their lifetimes. In Green's SNR catalog (Green 2019), 90% of Galactic SNRs are detected in the radio band, and only a small fraction of them appear faint in radio but relatively bright in the X-ray or optical bands. However, identifying Galactic SNRs in the radio band can be difficult due to the contamination from diffuse background radiation and the confusion with thermal radio radiation from H II regions. SNRs emit nonthermal synchrotron radiation, while H II regions emit thermal free–free radiation. These two types of radiation can be distinguished by their spectral indices. The spectral index $\alpha$ is defined by $S \propto \nu^{\alpha}$, where $S$ is the flux density and $\nu$ is the frequency. For synchrotron radiation, $\alpha \sim -0.5$, while for free–free radiation, $\alpha \sim 0$. Additionally, the polarized radio radiation, metal-rich or under-ionized X-ray plasma, and a low mid-infrared to radio flux ratio can all serve as evidence for identifying an SNR (Anderson et al. 2017).

Several radio surveys have been carried out in the past two decades to search for new Galactic SNRs. A survey by Brogan et al. (2006) using the 330 MHz multiconfiguration Very Large Array (VLA) led to the discovery of 35 new SNRs. The 1.4 GHz Multi-Array Galactic Plane Imaging Survey (MAGPIS) identified 49 SNR candidates (Helfand et al. 2006). Another survey, the Second Epoch Molonglo Galactic Plane Survey at 843MHz, resulted in the detection of 23 new SNR candidates (Green et al. 2014). Anderson et al. (2017) used The HI, OH, Recombination line survey of the Milky Way (THOR) and lower-resolution VLA 1.4 GHz Galactic Plane Survey (VGPS) continuum data to locate 76 new SNR candidates. These candidates were primarily identified by their shell-like morphology and the absence of mid-infrared (MIR) radiation.

Based on the spectral index and polarization, some of the candidates have been confirmed as SNRs using GaLactic and Extragalactic All-sky Murchison Widefield Array (GLEAM) survey at 80–300 MHz (Hurley-Walker et al. 2019a) and the 4–8 GHz global view on star formation (GLOSTAR) survey (Dokara et al. 2023). Besides the surveys using radio interferometric telescopes, single-dish radio observations have also helped in identifying SNRs, especially for those with large angular sizes (Gao et al. 2011, 2022).

Recent radio surveys greatly enlarged the population of SNR candidates, but confirming these Galactic SNRs has been difficult. To increase the sample of Galactic SNRs and understand what hampers us from finding new SNRs in the radio band, a detailed study of these candidates and their environments is important. A large fraction of the 76 new SNR candidates discovered by THOR +VGPS continuum data (Anderson et al. 2017) are distributed near $l = 28°.5$. This indicates intense stellar activity could exist within this region. In this paper, we use GLEAM and THOR +VGPS data to derive a spectral index map covering the region $26°.6 < l < 30°.6$, $|b| \leqslant 1°.25$ at 118–1420 MHz. By using the spectral index map, we identify several candidates as SNRs and further analyze the variation of their spectral index across different frequency ranges. Additionally, we found a molecular superbubble spatially situated in this region and investigated a long H$\alpha$ filament projected near pulsar wind nebula (PWN) candidate G29.38+0.10, We also discussed the association of pulsars with the newly identified SNRs, and the potential factors contributing to the missing Galactic SNRs.

## 2. Data and Catalogs

We retrieved data in four frequency bands centered at 1.4 GHz, 200 MHz, 154 MHz, and 118 MHz from THOR +VGPS and GLEAM surveys. The Green's SNR catalog (Green 2019, 2022) of known SNRs, the SNR candidates reported by Anderson et al. (2017), and the Wide-Field Infrared Survey Explorer (WISE) Catalog of Galactic H II regions (Anderson et al. 2014) are also used in this work.

### 2.1. THOR+VGPS

THOR is a survey of atomic, molecular, and ionic gas radiation in the northern Galactic plane, using the C-configuration of the VLA to observe 21 cm H I lines, four OH lines, and nineteen radio recombination lines, and a fully polarized continuous emission spectrum with a frequency of 1–2 GHz, observed with a resolution of $\sim 20''$ in the Galactic plane from $14°.5 < l < 67°.4$, $|b| \leqslant 1°.25$ (Beuther et al. 2016; Wang et al. 2020). To tackle the short-spacing issue, the THOR continuum data are combined with VGPS continuum data using the task "feather" in CASA software package (Wang et al. 2018). To avoid the missing flux problem, the VGPS data with an angular resolution of $60''$ are produced by combining the VLA D-configuration interferometric data and the single-dish observations from the Green Bank Telescope and the 100 m Effelsberg radio telescope (Stil et al. 2006). Hence, the combination of THOR data and VGPS data can provide radio images with a good angular resolution, but without losing the flux of large-scale structures. The resulting data set is called "THOR+VGPS[5]" with a resolution of $\sim 25''$ due to smoothing (Wang et al. 2018).

---

[5] https://thorserver.mpia.de/thor/image-server/

THOR+VGPS data in $26°.6 < l < 30°.6$, $|b| \leqslant 1°.25$ with central frequency 1.4 GHz are used in this work. The noise of THOR+VGPS data is dominated by side-lobe noise, and the noise maps were estimated by the averaged residual image from the clean process (Wang et al. 2018).

### 2.2. GLEAM

The GLEAM survey is a project that observed half of the observable Galactic plane using the Murchison Widefield Array between the frequency 72 and 231 MHz with an angular resolution $4'-2'$ (Hurley-Walker et al. 2016, 2019b). The currently released data cover the sky range of $345° < l < 67°$, $180° < l < 240°$, $1° \leqslant |b| \leqslant 10°$ for the center of the Milky Way, and $|b| \leqslant 10°$ for other regions (Hurley-Walker et al. 2019b). GLEAM used the Phase I configuration with the shortest baseline of 7.72 m, allowing maximum angular scales of 118 MHz, 154 MHz, and 200 MHz to reach $\sim 19°$, $\sim 14°$, and $\sim 11°$, respectively. The sources of this work are all smaller than the maximum angular scales, so we do not expect a strong missing flux problem for these sources.

The wideband GLEAM data with frequencies 103–134 MHz, 139–170 MHz, and 170–231 MHz are used together with the THOR+VGPS data to derive the spectral index of SNRs and SNR candidates. The 72–80 MHz GLEAM data are not used due to their low angular resolution. Noise maps of GLEAM data are estimated by BANE (Hancock et al. 2018) and can be overestimated in Galactic Plane regions due to the algorithm of BANE.

### 2.3. Catalogs

The Green's SNR catalog G19 (Green 2019) contains 294 known Galactic SNRs by the end of 2018, with the information of coordinates, angular size, type, flux density, and spectral index. Among them, five known supernova remnants named G27.4+0.0 (Kes 73), G28.6−0.1, G29.6+0.1, and G29.7−0.3 (Kes 75) are located in $26°.6 < l < 30°.6$, $|b| \leqslant 1°.25$. The most up-to-date Green catalog is the 2022 December version released online (Green 2022). The number of Galactic SNRs reaches 303 in this version and two new SNRs named G28.3 +0.2 and G28.7−0.4 in the region of this work are added. However, Green's SNR catalog is still not complete, with more SNRs to be added in further radio studies.

Anderson et al. (2017) used THOR+VGPS data to discover 76 new SNR candidates (hereafter, A17 SNR candidates). 19 of them fall in our interested sky region, where two of these candidates (G28.36+0.21 and G28.78−0.44) have already been included in the Green's SNR Catalog 2022. However, these two SNRs are newly identified, and their spectral indices are still denoted as uncertain, so they are recognized as SNR candidates for further study in this work.

We also used the WISE Catalog of Galactic H II regions (Anderson et al. 2014), which contains $\sim 8000$ Galactic H II regions and H II region candidates. These sources were identified through the characteristic MIR morphology of H II regions (Wright et al. 2010).

The distribution of G19 SNRs, A17 SNR candidates, and WISE H II regions on the 1.4 GHz THOR+VGPS data in $26°.6 < l < 30°.6$, $|b| \leqslant 1°.25$ is shown in Figure 1.

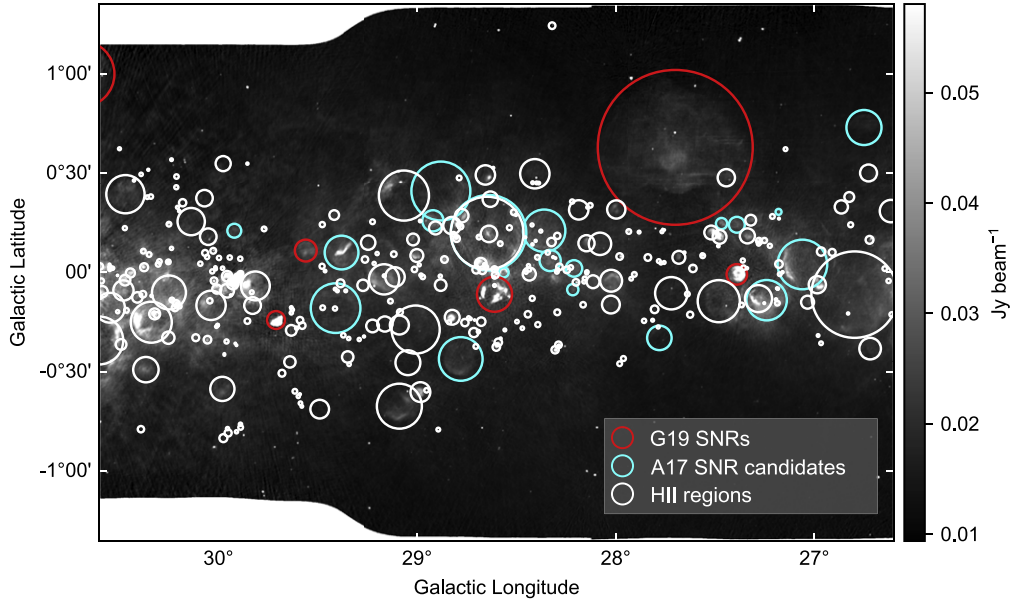

**Figure 1.** THOR+VGPS 21 cm continuum data in $26°.6 < l < 30°.6$, $|b| \leqslant 1°.25$. The known SNRs in the Green's SNR catalog (Green 2019; G19 SNRs) are enclosed by red circles, The SNR candidates discovered by Anderson et al. (2017; A17 SNR candidates) are enclosed by cyan circles, and WISE known and candidate H II regions are enclosed by white circles.

## 3. Results

### 3.1. Spectral Index Map

We first provide a spectral index map of the large sky area in $26°.6 < l < 30°.6$, $|b| \leqslant 1°.25$, by using the flux density maps of 1.4 GHz THOR+VGPS continuum data and three wideband GLEAM data. The comparison requires the pixels of four-wave-band data to be aligned and the spatial resolution of these data to be the same. First, these data are resampled to the same coordinates and pixel grid using flux-conserving spherical polygon intersection. This method projects both input and output pixels onto the celestial sphere, calculates the area of spherical polygon intersection between two pixels, and obtains output pixels through area-weighted averaging. Then, all images are smoothed to the same angular resolution of $4'$. These images are further used for spectral index maps and spectral analysis.

After converting the unit of flux density from Jy beam$^{-1}$ to Jy pixel$^{-1}$, we used the least squares fitting method to fit the four-band data pixel by pixel with a power-law spectrum, $S_\nu \propto \nu^\alpha$, and obtain the spectral index $\alpha$. The uncertainty map $\sigma(i, j)$ is determined by the propagation of the noise map of the original data. During the smoothing process, the uncertainty propagation is done according to the method by Klein (2021).

Diffuse radiation in the Galactic plane can contaminate individual sources by contributing extra emissions and thus change the spectral index of the sources. Thus, we remove the large-scale background from these data using the FINDBACK command from the Starlink software package (Currie et al. 2014; Berry et al. 2022). This algorithm twice filters the data (first to find the minimum and second to find the maximum) with a rectangular kernel to obtain the baseline component. The baseline data are then smoothed to minimize sharp edges and corrected to avoid an underestimation of the true background. Finally, the obtained background is subtracted from the input data to give the final image. The kernel size is specified as the size of the smallest features that are to remain in the final background-subtracted images. Therefore, we took a kernel size of $24.1'$, which is similar to the size of the largest SNR G27.8+0.6 in the interested field.

Figure 2 shows the $6\sigma$ spectral index maps before and after background removal, in which we only use pixels with flux density values 6 times above the uncertainty. It can be seen that the spectra are steeper after removing the background. Before the background removal, only three small-scale SNRs (red solid circles) in the field show spectral indices $\alpha < -0.2$, and all the SNR candidates (dashed circles and cyan circles) and a large-scale SNR (G27.8 +0.6) have flat spectra comparable to that of the background area. From the background-removed image, we found more sources likely to have a nonthermal origin. These sources are either faint or extended so that they suffer stronger contamination from the diffuse background. Hence, it is effective to reduce the contamination of diffusive radiation and obtain a more reliable spectral index map of nonthermal sources by removing the background.

As shown in the background-removed spectral index map, all the G19 SNRs have strong evidence of nonthermal emission, except for G29.6+0.1. We found steep spectra with $\alpha < -0.2$ from four A17 SNR candidates, namely G26.75+0.73, G27.06 +0.04, G28.36+0.21, and G28.78−0.44 (enclosed by red dashed circles). Besides SNRs, several point sources with nonthermal emissions are identified and denoted as orange × signs in Figure 2.

### 3.2. Spectral Analysis of Some SNR Candidates

Hereafter, we calculate the integrated flux densities and spectral indices of the five A17 SNR candidates using aperture photometry in four radio bands. Besides the four A17 SNR candidates which show clear steep spectra, G29.38+0.10 is also included despite having a flat spectrum, since it was considered as PWN in previous studies (Hurley-Walker et al. 2019a; Dokara et al. 2021, 2023). Unlike SNRs, PWNe have a flat radio spectral index of $\alpha \approx 0-{-}0.3$ (Reynolds et al. 2017), which is assumed attributed to the acceleration through magnetic-field reconnection (Xu et al. 2019).

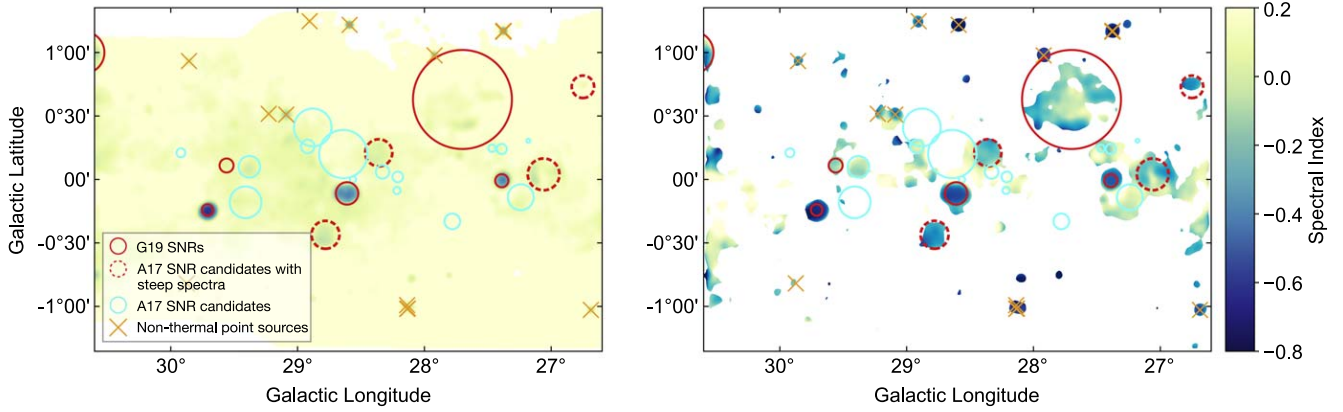

**Figure 2.** The 6 σ spectral index maps overlayed with G19 SNRs (red circles), A17 SNR candidates with steep spectra (red dashed circles), the rest A17 SNR candidates (cyan circles), and the point sources that may correspond to nonthermal radiation (orange × signs) in $26°.6 < l < 30°.6$, $|b| \leqslant 1°.25$. Left: the spectral index map derived from data before removing the background radiation. Right: the spectral index map derived from data after removing the background radiation and only the parts that exceed three times the average flux density of the clean region are shown to highlight the significant sources. The color bar for the spectral index is the same for both the left and right panels.

We calculate the integrated spectral index of SNR candidates using the four-band radio maps. The source spectrum is selected from a circular region with an angular radius of $\theta = \sqrt{\theta_{\rm src}^2 + \theta_{\rm kernel}^2}$, where $\theta_{\rm src}$ and $\theta_{\rm kernel}$ are the source radius and smoothing kernel size, respectively, at 1.4 GHz. A source-free concentric ring with a width of about two times the source radius is chosen to subtract the background. If other sources such as H II region fall within the source aperture or background ring, we adjust the shape and border of the source aperture or background ring to exclude the unrelated sources. The integrated flux densities are then computed as

$$I = I_s - I_b \frac{n_s}{n_b}, \quad (1)$$

where $I$ is the net source flux densities by removing the background level from the background region, $I_s$ is the measured integrated source flux densities, $I_b$ is the background flux densities, $n_s$ and $n_b$ are the numbers of pixels in the source and background region, respectively. We assume the background level of each source to be constant (i.e., the uncertainty that comes from the fluctuation of background level is not considered). In that case, the uncertainty from the noise map is so small that it can be ignored. Then the uncertainty is determined by the absolute intensity calibration ∼20% for the THOR+VGPS data (Anderson et al. 2017) and the external flux density scale error ∼13% for the GLEAM data (Hurley-Walker et al. 2019b).

The radio flux densities and spectral indices of five A17 SNR candidates are listed in Table 1, which also give the source coordinates, radius, fitted 1.0 GHz flux density, and the radio morphological type.

We noticed that the spectral index of an SNR can vary with frequency due to, for example, absorption along the line of sight. Hence, we use a figure for each of these five A17 SNR candidates to compare the spectral index derived in this work and other works between different frequency intervals. The THOR+VGPS continuum images, the flux density spectrum using THOR+VGPS and GLEAM data, and the spectral index comparison between different frequency intervals of the five A17 SNR candidates are shown in Figure 3.

G26.75+0.73 shows a likely composite morphology in THOR+VGPS data and has a faint emission in the GLEAM data. We derived a nonthermal spectral index of −0.41 ± 0.10, consistent with the value of ∼ −0.4 that Dokara et al. (2021) measured using 200MHz GLEAM data and 1.4 GHz THOR+VGPS data. Together with its high degree of overestimated polarization of ∼0.70 observed in the GLOSTAR-VLA (Dokara et al. 2021), we confirm it as an SNR.

G27.06+0.04 is a shell-type SNR candidate first detected in the MAGPIS survey (Helfand et al. 2006). The southwest region of G27.06+0.04 overlaps with a bright H II regions (enclosed by a white circle in Figure 3), which should be excluded to obtain the source flux density. We derive an overall spectral index of −0.23 ± 0.10. Dokara et al. (2018) measured an average spectral index of −0.53 ± 0.22 for the eastern arc of G27.06+0.04 from the TGSS-NVSS spectral index map (150 MHz–1.4 GHz; de Gasperin et al. 2018). We measure its overall flux density of 4.29 ± 0.86 Jy at 1.4 GHz with the THOR+VGPS data, larger than 1.4 ± 0.3 Jy measured by Dokara et al. (2021) for only the arc. Therefore, our overall spectral index is flatter than the value of −0.65 ± 0.31 from (Dokara et al. 2021) for the arc. Its nonthermal spectrum supports it as an SNR.

G28.36+0.21 has a circular shell structure which is typical of SNRs. We derive a spectral index of −0.20 ± 0.10. Hurley-Walker et al. (2019a) derived a steep spectral index of −0.72 ± 0.10 within GLEAM data from 70–230 MHz. Dokara et al. (2023) obtained a spectral index of −0.28 ± 0.11 using GLEAM 200 MHz, THOR+VGPS, and GLOSTAR, consistent with the value we derived in this work. The spectral index prefers a nonthermal nature. Combined with its structure, it is a shell-type SNR with a high confidence level. The fraction polarization of ∼2% from the GLOSTAR survey (Dokara et al. 2023) gives further evidence of it as an SNR.

G28.78−0.44 was identified as an SNR candidate in the MAGPIS (Helfand et al. 2006), THOR (Anderson et al. 2017), GLEAM (Hurley-Walker et al. 2019a), and GLOSTAR (Dokara et al. 2023) surveys. In this work, we obtain a spectral index of −0.38 ± 0.10. Hurley-Walker et al. (2019a) derived a spectral index of −0.79 ± 0.12 and Dokara et al. (2021) calculated an average spectral index of −0.75 ± 0.22 for a part of the shell with the TGSS-NVSS spectral index map (de Gasperin et al. 2018). The fractional polarization measured by

Table 1
A17 SNR Candidates

| Name | GLong (°) | GLat (°) | Radius[a] (arcmin) | $S_{1.4\,G}$ (Jy) | $S_{200\,M}$ (Jy) | $S_{154\,M}$ (Jy) | $S_{118\,M}$ (Jy) | $\alpha$ | $S_{1.0\,G}$[b] (Jy) | $R^2$[c] | Type[d] | Ref. |
|---|---|---|---|---|---|---|---|---|---|---|---|---|
| G26.75+0.73 | 26.750 | 0.730 | 5.3 | 0.53 ± 0.11 | 1.16 ± 0.15 | 1.24 ± 0.17 | 1.51 ± 0.20 | −0.41 ± 0.10 | 0.60 ± 0.10 | 0.95 | C? | 1[f, g] |
| G27.06+0.04 | 27.060 | 0.040 | 7.5 | 4.29 ± 0.86 | 6.08 ± 0.79 | 7.31 ± 0.95 | 7.50 ± 0.98 | −0.23 ± 0.10 | 4.52 ± 0.76 | 0.93 | S | 1[f, g], 2[f] |
| G28.36+0.21[e] | 28.360 | 0.210 | 6.4 | 2.92 ± 0.59 | 4.05 ± 0.53 | 4.76 ± 0.62 | 4.75 ± 0.62 | −0.20 ± 0.10 | 3.09 ± 0.52 | 0.94 | S | 3[f], 4[f, g] |
| G28.78−0.44[e] | 28.780 | −0.436 | 6.6 | 2.68 ± 0.54 | 5.28 ± 0.69 | 6.44 ± 0.84 | 6.77 ± 0.88 | −0.38 ± 0.10 | 3.01 ± 0.50 | 0.96 | S | 1[f, g], 3[f], 4[f, g] |
| G29.38+0.10[h] | 29.380 | 0.100 | 5.1 | 2.75 ± 0.55 | 3.22 ± 0.42 | 3.47 ± 0.46 | 3.06 ± 0.40 | −0.06 ± 0.09 | 2.88 ± 0.46 | 0.46 | C? | 1[f, g], 3[f], 4[f, g] |

**Notes.**
**References.** (1) Dokara et al. (2021); (2) Dokara et al. (2018); (3) Hurley-Walker et al. (2019a); (4) Dokara et al. (2023).
[a] The radius defined in (Anderson et al. 2017) using THOR+VGPS data.
[b] The 1.0 GHz flux density is obtained by fitting.
[c] The Goodness of fit is defined by $R^2 = 1 - \frac{\sum_{n=1}^{4}(S_i - \widehat{S_i})^2}{\sum_{n=1}^{4}(S_i - \bar{S})^2}$, where $S_i$ is the measured flux density of $i$th band, $\widehat{S_i}$ is the fitted flux density of $i$th band, and $\bar{S}$ is the mean flux density of four bands.
[d] "S" for shell-type, "C" for composite. Question marks ("?") indicate uncertainty in the classification.
[e] Identified as an SNR in the Green catalog 2022 December version.
[f] This paper confirms the source as an SNR using spectral index measurement, but the radio bands are different from this work.
[g] This paper confirms the source as an SNR using polarization measurement.
[h] The flux densities and spectral index of G29.38+0.10 are contaminated by a background galaxy candidate.

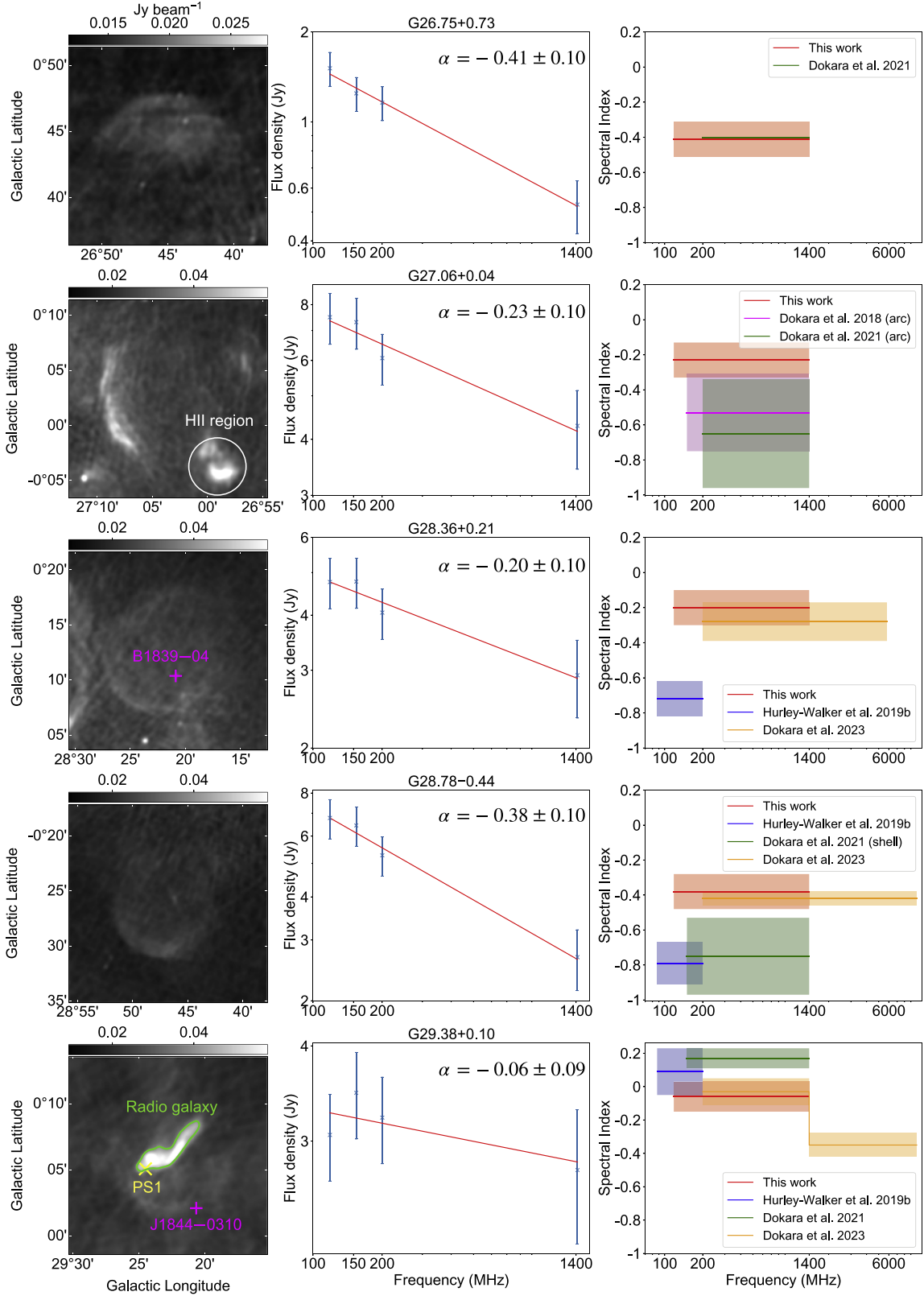

**Figure 3.** From top to bottom: G26.75+0.73, G27.06+0.04, G28.36+0.21, G28.78−0.44, and G29.38+0.10. Left: the THOR+VGPS continuum images of the five A17 candidates. The white circle encloses a H II that overlaps with G27.06+0.04. The magenta cross signs denote the two known pulsars spatially located within G28.36+0.21 and G29.38+0.10. The green contour shows a background radio galaxy candidate (Castelletti et al. 2017). The putative pulsar of G29.39+0.10 is marked as a yellow × sign. Middle: The four wave bands' flux density spectrum of the five A17 candidates. The blue × signs denote the flux density measured in 118 MHz, 154 MHz, and 200 MHz using GLEAM data and 1.4 GHz using THOR+VGPS continuum data, with 1-$\sigma$ error bars. The red line is the best-fit line. Right: the spectral indices of the five A17 candidates between different frequency intervals derived by this work and previous works. The width of the colored bands represents the error in the spectral index. For G27.06+0.04, Dokara et al. (2018) and Dokara et al. (2021) derive the spectral index for the eastern arc. For G28.78−0.44, Dokara et al. (2021) derive the spectral index for a part of the shell.

(Dokara et al. 2023) is ∼4%. A more broadband spectral index of −0.42 ± 0.04 derived by Dokara et al. (2023) is consistent with our value. The polarized emission of the shell and the nonthermal spectrum give strong evidence for an SNR.

G29.38+0.10 consists of a bright inner S-shaped structure and a faint outer shell, as shown in the THOR+VGPS continuum image. We derive a spectral index of −0.06 ± 0.09 in the frequency range 118–1400 MHz, which is typical for a PWN ($\alpha \approx 0$–0.3; Reynolds et al. 2017) but overlaps that of a thermal source. Hurley-Walker et al. (2019a) derived a spectral index of 0.09 ± 0.14 which is similar to the 0.17 ± 0.06 (Dokara et al. 2021) obtained from the TGSS-NVSS spectral index map (de Gasperin et al. 2018). Dokara et al. (2023) derived a value of −0.03 ± 0.08 in 200–1400 MHz, which is consistent with this work. In the higher frequency of 1.4–10 GHz, Dokara et al. (2023) obtained a spectral index of the whole source is −0.35 ± 0.07. This spectral break is also a typical feature of PWNe (Pacini & Salvati 1973; Reynolds & Chevalier 1984; Xu et al. 2019). However, the radio flux of the source is dominated by a bright S-shaped structure in the interior (marked as a green contour in Figure 3), which is considered to be a background radio galaxy candidate (Castelletti et al. 2017). Therefore, the measured spectral index of G29.38+0.10 is contaminated by the extragalactic source, and the spectral index after removal of the radio galaxy is not given in this paper. For the whole region, the degree of polarization is ∼5.5% derived from the GLOSTAR survey (Dokara et al. 2023). Though the morphology of G29.38+0.10 agrees with a PWN inside an SNR shell (e.g., a composite SNR), we still leave it as an SNR candidate that waits for further observations and studies.

For each source, the spectral index obtained by subtracting the local background level (Figure 3) is consistent with the average value in the spectral index map (Figure 2), which subtracts the large-scale diffuse background. Compared to the TGSS-NVSS spectral index map (de Gasperin et al. 2018) that uses two-wave-band data at 150 MHz and 1.4 GHz, we use four-wave-band data from 118 MHz to 1.4 GHz to give a better coverage of the middle frequency. Furthermore, the frequency range used in this work is broader than 70–200 MHz adopted by Hurley-Walker et al. (2019a) and extends to lower frequencies than that used by Dokara et al. (2023). Therefore, our work provides a useful complement to previous radio studies and allows for a cross check.

## 4. Discussion

### 4.1. Spatial Correspondence of SNRs/SNR Candidates with an Expanding Molecular Superbubble

There are numerous SNRs/SNR candidates distributed in the region $26°.6 < l < 30°.6$, $|b| \leqslant 1°.25$. To understand whether this is just a coincidence of projection or whether some of these SNRs/SNR candidates are born close to each other, we searched for potential spatial correspondence between these SNRs/SNR candidates and molecular clouds (MCs). We use the $^{12}$CO and $^{13}$CO $J = 1$–0 data from FOREST Unbiased Galactic plane Imaging survey with the Nobeyama 45 m telescope (FUGIN; Umemoto et al. 2017) to perform investigation.

We considered CO observations in the range $26°.6 < l < 30°.6$ to look for molecular gas. We found an expanding bubble with a radius of ∼0°.75 at $V_{LSR} = 79.7$ km s$^{-1}$, as shown in the top panel of Figure 4. The bubble is centered at $l=28°.8$, $b=0°.12$. Interestingly, several known SNRs and newly identified SNRs (red circles) spatially correspond to this molecular structure. Based on existing data, we have not found clear evidence of shock-cloud interaction (e.g., $^{12}$CO $J = 1$-0 line broadening), so the physical association between the SNRs and clouds needs further data for investigation. The bottom panel of Figure 4 illustrates the position–velocity (PV) diagram at $l = 29°.1$ and $-1° < b < 1°$ cutting from $(l, b) = (29°.1, -1°)$ to $(29°.1, 1°)$. The presence of a half-annular pattern at the velocity interval $V_{LSR} = 80$–100 km s$^{-1}$ in the PV diagram reveals that the molecular structure is expanding toward the side away from us (redshift) with an expansion velocity of ∼20 km s$^{-1}$. The systemic velocity of the molecular structure is probably at ∼80 km s$^{-1}$, where the molecular shell is most clearly seen (top panel of Figure 4), and the PV diagram shows a flat pattern indicative of unperturbed gas. This expansion pattern could be attributed to an expanding molecular bubble due to a collision of winds with the dense ISM. The lack of blueshifted CO emission suggests that the molecular gas is mostly distributed in the backside.

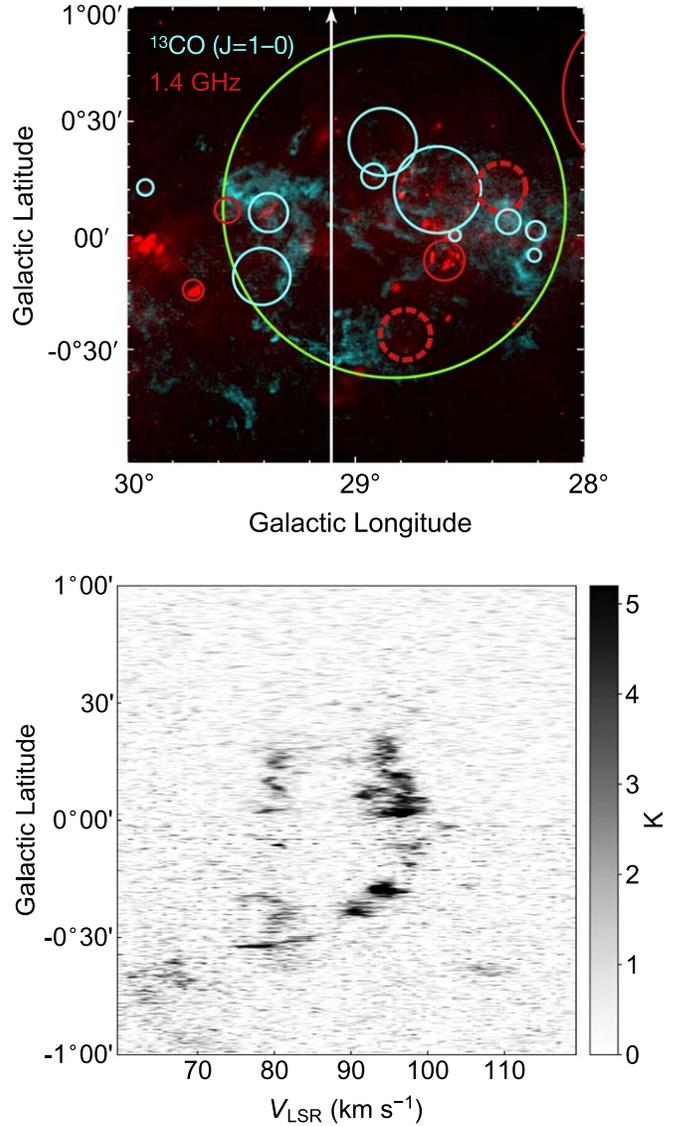

**Figure 4.** Top: the THOR+VGPS continuum image (red) and the $^{13}$CO ($J = 1$–0) 79.7 km s$^{-1}$ image (cyan), overlayed with G19 SNRs (red circles), newly confirmed SNRs (red dashed circles), and the A17 SNR candidates except for the four newly confirmed SNRs (cyan circles). The molecular bubble is enclosed by a green circle. Bottom: the PV diagram along the white line in the top panel at $l = 29°.1$ and $-1° < b < 1°$.

To estimate the distance of the molecular bubble, we use updated solar motion parameters provided by Reid et al. (2014) to correct the measured $V_{LSR}$ and then use a Monte Carlo method (Wenger et al. 2018) to derive kinematic distances and uncertainties. With the systematic $V_{LSR}$ of 80 km s$^{-1}$, we derive a near kinematic distance of $4.7^{+0.5}_{-0.4}$ kpc and a far kinematic distance of $10.0^{+0.3}_{-0.6}$ kpc. To determine whether the bubble is located in the near or far kinematic distance (i.e., kinematic distance ambiguity), we make use of the Galactic Ring Survey Molecular Cloud identification (GRSMC) catalog (Roman-Duval et al. 2009). The kinematic distance ambiguity of MCs in this catalog is resolved based on the continuum absorption or self-absorption of the H I 21 cm line (Roman-Duval et al. 2009). According to the GRSMC catalog, the kinematic distance of MCs near the location of this molecular bubble, with a $V_{LSR}$ of ∼80 km s$^{-1}$, is close to 4.7 kpc. Therefore, we adopt the near kinematic distance of ∼4.7 kpc for further analysis of this bubble.

The angular radius of the bubble is 0°.75, which is defined by the curvature of the bright southeast rim. At the distance of ∼4.7 kpc, the radius of the bubble is $R \sim 62$ pc, and the mass of the molecular gas in $V_{LSR} = 70$–100 km s$^{-1}$ is obtained as $M_{H_2} \sim 2 \times 10^6\, M_\odot$. The total kinetic energy of the expanding molecular bubble is thus $E_k = 1/2 M_{H_2} dV^2 \sim 8.0 \times 10^{51}$ erg, where the expansion velocity $dV$ is ∼20 km s$^{-1}$. The bubble size and the large kinetic energy cannot be produced by a single star or a normal supernova explosion, but are consistent with the superbubbles (typically >30 pc and ∼$10^{52}$ erg, respectively; Watkins et al. 2023). A Superbubble is produced by the combined effect of stellar winds and SNe from tens of massive stars in OB associations and consists of a large shell of cold gas and a hot interior (Mac Low & McCray 1988; Yadav et al. 2017). The shells of some young superbubbles can be traced by molecular gas, where shock-cloud interaction occurs (e.g., 30 Dor C; Sano et al. 2017; Yamane et al. 2021). The extinction on the Galactic plane may prevent us from observing OB stars, but the molecular superbubble produced by these OB stars can provide a useful way to identify them.

Watkins et al. (2023) found 325 molecular superbubbles in the nearby galaxies using PHANGS–ALMA $^{12}$CO ($J = 2 - 1$) survey. They found an average injection efficiency (the ratio of the total kinetic energy $E_k$ of the molecular bubble to the total mechanical energy output by its stellar population $E$) $\eta \sim 10\%$ and an expansion parameter of $m \sim 0.25$ (bubble radius $R \propto t^m$). Adopting these values, we derive that the total mechanical energy from the stellar population inside the bubble is $E = E_k/\eta \sim 8.0 \times 10^{52}$ erg and the age of the bubble is $t = m^* R/dV \sim 0.8$ Myr.

These molecular superbubbles identified by Watkins et al. (2023) have an age range of 0.7–7.5 Myr and expansion velocity range of ∼2–25 km s$^{-1}$, which are consistent with our estimated properties of the molecular bubble at around $l = 29°$ (∼0.8 Myr and ∼20 km s$^{-1}$, respectively). According to the discussion above, we suggest that the molecular bubble we found is part of a young superbubble with mechanical energy of $8.0 \times 10^{52}$ erg.

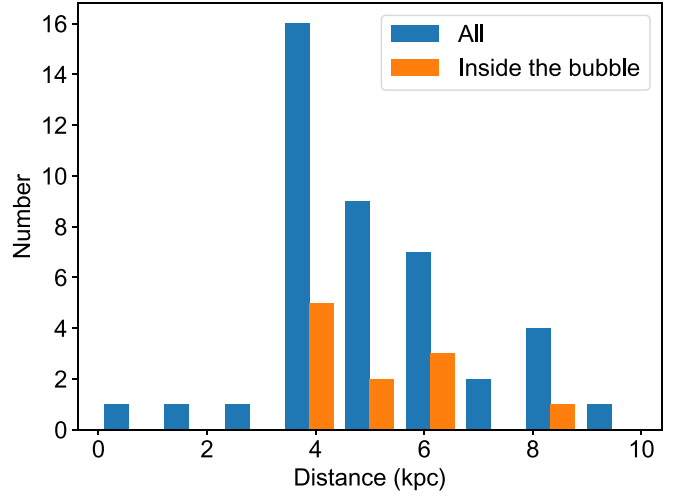

**Figure 5.** Blue: the histogram of distances of all known pulsars within $26°.6 < l < 30°.6$, $|b| \leqslant 1°.25$. Orange: the histogram of distances of known pulsars within the sky area covered by the molecular bubble, which is denoted as a green circle in Figure 4.

### 4.2. Pulsars Associated with SNR Candidates

A core-collapse SN explosion eventually forms a compact object, either a neutron star or a black hole. One third of Galactic SNRs are found to be associated with neutron stars, which helps to establish the core-collapse origin of the SNRs and confirm the SNRs (Ferrand & Safi-Harb 2012). Due to the asymmetry of some SN explosions, neutron stars can get a kick velocity and move outside of their host SNRs as they age. However, for young SNRs that expand fast, neutron stars can still sit within or near their host SNRs, so finding pulsars that are associated with SNR candidates can be further evidence to identify SNRs. We found 44 known pulsars within $26°.6 < l < 30°.6$, $|b| \leqslant 1°.25$ from the ATNF Pulsar Catalogue and listed information of these pulsars in Table 2 in Appendix A. 13 of these pulsars are spatially located inside the region of the molecular bubble, which is denoted as a green circle in Figure 4. Figure 5 shows that most pulsars are located at a distance of ∼4–6 kpc, which is consistent with the estimated distance of the molecular superbubble ($4.7^{+0.5}_{-0.4}$ kpc). This indicates a possible association between many of the pulsars and the molecular superbubble.

There are two known pulsars, B1839−04 and J1844−0310, spatially located within the SNR G28.36+0.21 and SNR candidate G29.38+0.10, respectively (marked as cross signs in Figure 3). The dispersion measure (DM) -inferred distances of these two pulsars are ∼3.7 kpc and ∼6.0 kpc, with a large relative error (<90%; Yao et al. 2017). Their characteristic ages estimated from the periods and period derivatives are ∼60 Myr and ∼1 Myr, respectively, which are much older than the typical SNR age of $\lesssim 10^5$–$10^6$ yr. However, it is noteworthy that the characteristic age of a pulsar can highly deviate from its true age (Kaspi et al. 2001). The good spatial correspondence hints that B1839−04 could be the possible pulsar of SNR G28.36+0.21. For G29.39+0.10, an X-ray point source named CXO J18443.4−030520 (hereafter, PS1) has been found in the "head" of the PWN candidate and thus proposed as the associated neutron star (Castelletti et al. 2017; Petriella 2019, marked as yellow × sign in Figure 3), while J1844−0310 is situated outside the region of the PWN candidate. However, the relation between the X-ray point source and G29.39+0.10 has not been established due to the nondetection of pulsations. Further studies are needed to find out whether J1844−0310 or PS1 is the related neutron star.

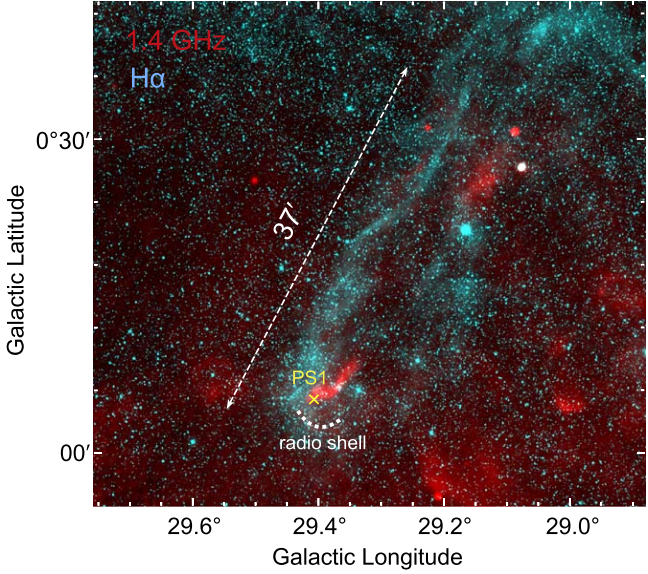

**Figure 6.** The THOR+VGPS continuum image (red) and the SuperCOSMOS Hα image (cyan). A yellow × sign denotes the potential pulsar of G29.38 +0.10. The white dashed line shows the length of the Hα structure. The white dotted line shows the approximate position of the radio shell.

### 4.3. A Hα Filament Projected near PWN Candidate G29.38+0.10

Using the SuperCOSMOS Hα Survey (Parker et al. 2005), we found a long Hα filament structure extending from the SNR/PWN candidate G29.38+0.10 to ∼37′ the upper-right (see Figure 6). The bottom-left end of the Hα filament is curved and spatially matches the radio shell of G29.38+0.10.

Since some pulsars can move quickly at speed over several hundred km s$^{-1}$ and form a comet-shaped PWN (Kargaltsev et al. 2017), it is worthwhile to explore whether this long Hα filament is a bow-shock tail structure produced by G29.38+0.1 as it quickly runs to the southwest. Here we assume the compact X-ray source PS1 (see yellow × sign in Figure 6) is the pulsar of G29.38+0.1.

Fast-moving PWN can ionize the surrounding medium and excite Hα emission (Brownsberger & Romani 2014). As the PWN passed the ISM, the Hydrogen atoms lose the ionization source and thus recombine with a characteristic timescale (Draine 2011):

$$\tau_{\rm rec} = \frac{1}{\alpha(T) n_{\rm ISM}} \approx 1.2 \times 10^5 \left(\frac{n_{\rm ISM}}{1 \text{ cm}^{-3}}\right)^{-1} \text{yr}, \quad (2)$$

where $\alpha(T) \approx 2.5 \times 10^{-3}$ is the recombination rate for case B (optically thick case) at a typical temperature of $10^4$ K, and the ISM is assumed to be fully ionized with a density $n_{\rm ISM}$. The partially ionized medium would require a longer recombination time. If the Hα filament is the trail of the PWN at the distance $d = 6.5$ kpc (Petriella 2019), the length of the Hα filament is $l \sim 70 \left(\frac{d}{6.5 \text{ kpc}}\right)$ pc. Despite the large uncertainty in the distance, the length of the Hα filament is still two orders of magnitude longer than the known Hα tails produced by PWNe (Brownsberger & Romani 2014; Kargaltsev et al. 2017). The length of the Hα filament can be used to estimate the transverse velocity of the PWN:

$$v \sim l/\tau_{\rm rec} \sim 571 \left(\frac{d}{6.5 \text{ kpc}}\right) \left(\frac{n_{\rm ISM}}{1 \text{ cm}^{-3}}\right) \text{km s}^{-1}. \quad (3)$$

The transverse velocity $v$ scales with the ambient density. For $n_{\rm ISM} = 1$ cm$^{-3}$, $v$ is larger than typical pulsars but reasonable for a fast-moving pulsar. However, we need to investigate if this velocity is compatible with other properties, such as the angular scale between the bow shock apex and the pulsar. The spin-down power of the PWN candidate has been estimated as $\dot{E} \sim 1 \times 10^{37}$ erg s$^{-1}$ based on the empirical relations between $\dot{E}$ with the PWN luminosity (Petriella 2019). The predicted standoff angle between the pulsar and the contact discontinuity, where the pulsar wind and ISM ram pressure balance, is $\theta_0 = (\dot{E}/4\pi c \rho_{\rm ISM} v^2)^{1/2}/d$, with the gas density $\rho_{\rm ISM} = 1.37 m_p n_{\rm ISM}$ and the proton mass $m_p$. The angular scale between the bow shock apex and the pulsar can be estimated as (Chen et al. 1996; Bucciantini & Bandiera 2001; Brownsberger & Romani 2014):

$$\theta_a = 1.3\theta_0$$
$$= 0.80'' \left(\frac{\dot{E}}{10^{37} \text{ erg s}^{-1}}\right)^{\frac{1}{2}}$$
$$\times \left(\frac{d}{6.5 \text{ kpc}}\right)^{-2} \left(\frac{n_{\rm ISM}}{1 \text{ cm}^{-3}}\right)^{-\frac{3}{2}}. \quad (4)$$

This angular scale corresponds to a distance scale of $0.025 \, (\dot{E}/10^{37} \text{ erg s}^{-1})^{1/2} (d/6.5 \text{ kpc})^{-1} (n_{\rm ISM}/1 \text{ cm}^{-3})^{-3/2}$ pc between the PWN bow shock and pulsar, similar to that of the currently known Hα pulsar bow shocks in the range of ∼0.01–1 pc (Brownsberger & Romani 2014; Kargaltsev et al. 2017). The measured distance between the southeastern apex of the Hα and PS1 is ∼4′, requiring a very high-density medium with $n_{\rm ISM} \sim 45$ cm$^{-3}$ to match the predicted value in Equation (4). However, such a high-density medium would infer an unrealistically high transverse velocity of the pulsar (see Equation (3)).

A problem with linking the Hα filament and PWN candidate is the high extinction in the inner Galactic plane. The PWN candidate is highly absorbed with a Hydrogen column density of $N_{\rm H} = 10.15 \times 10^{22}$ cm$^{-2}$ (Castelletti et al. 2017; Petriella 2019), which is converted to a large optical extinction $A_V \sim 35$ following the empirical $N_{\rm H}$–$A_V$ relation ($N_{\rm H} = 2.87 \times 10^{21} A_V$ cm$^{-2}$; Foight et al. 2016). The optical emission at the distance of G29.38+0.10 can be too heavily absorbed to be seen. Therefore, the Hα filament is likely only a foreground structure that coincides with G29.38+0.10 on the line of sight.

### 4.4. The Problem of Missing Galactic Supernova Remnants

Observation limitations could be one of the major reasons for the problem of missing Galactic SNRs. Out of the 18 A17 SNR candidates within the region of $26°.6 < l < 30°.6$, $|b| \leqslant 1°.25$, we only identify four of them as SNRs. The remaining A17 candidates either exhibit flat spectra, resembling H II regions, or are too small and faint to be detected in our spectral index map. The angular resolution of the lowest frequency data is 4′, while the radius of some candidates is less than 3′. Consequently, most of these candidates are not resolved in low-frequency data due to their small angular size. This also prevents us from deriving a reliable spectral index free of contamination from the nearby emission. Compared to the known SNRs, the four newly identified SNRs have relatively flatter spectra, which may be due to background thermal emission or the fact that older SNRs tend to have flatter spectra (Bozzetto et al. 2017). However, Ranasinghe & Leahy (2023) did not find such an age-spectral index relation. Since the older SNRs are usually very faint, there could be a large amount of them left to be discovered with more sensitive radio surveys.

In the future, the situation is expected to change with the advent of next-generation interferometers like the Square Kilometer Array Observatory (SKA). These instruments will offer phenomenal angular resolution and sensitivity, enabling us to overcome the current limitations and improve our ability to identify SNRs.

Furthermore, if an SN explodes within low-density environments such as a superbubble, the resulting SNR can appear extremely faint in radio since the low-density medium swept by the shock wave cannot emit sufficient radiation (Chen et al. 2000). Detecting and identifying these particular SNRs using radio observations can become difficult. Moreover, SNe with low explosion energies, such as electron capture SNe, can produce short-lifetime SNRs. The existence of these short-lived SNRs can lead to an overestimation of the estimated number of Galactic SNRs. All of the aforementioned factors could contribute to the problem of missing Galactic supernova remnants.

## 5. Conclusion

We have studied SNR candidates covering the region in $26\overset{\circ}{.}6 < l < 30\overset{\circ}{.}6$, $|b| \leqslant 1\overset{\circ}{.}25$ using the radio survey data from 103 MHz to 1.4 GHz, and investigated their environment using molecular and optical data. With this spectral index map and further spectral index analysis of the SNR candidates in this region, we concluded that four sources named G26.75+0.73, G27.06+0.04, G28.36+0.21, and G28.78−0.44 are SNRs. G29.38+0.10 has a characteristic spectral index of PWN but is still an uncertain SNR due to the confusion of a radio galaxy.

Many of the SNR candidates and SNRs are projected in the vicinity of an expanding molecular shell with a radius of 62 pc at a distance of ∼4.7 kpc. We suggest that this shell is swept up by a superbubble with an age of ∼0.8 Myr and mechanical energy from its stellar population of $\sim 8.0 \times 10^{52}$ erg. We note that a large fraction of pulsars in our studied region fall in the distance 4–6 kpc, consistent with that of the molecular shell and indicative of an association. The good spatial correspondence suggests that B1839−04 is the possible pulsar of the newly identified SNR G28.36+0.21. For G29.39+0.10, PS1 is proposed as the associated neutron star, but the relation is uncertain due to the nondetection of pulsations. We found a long Hα filament spatially corresponds to G29.38+0.10 and excluded it as a bow-shock tail structure of a PWN. The problem of missing Galactic SNRs may arise from not only observation limitations, but also the environments and the intrinsic SN explosion properties. Future sensitive and high-resolution radio surveys (e.g., with SKA) will help to better understand the missing SNR problem.

## Acknowledgments

We thank N. Hurley-Walker for providing the noise maps of GLEAM data. This publication also makes use of data from FUGIN, FOREST Unbiased Galactic plane Imaging survey with the Nobeyama 45 m telescope, a legacy project in the Nobeyama 45 m radio telescope. This work was supported by the ATNF Pulsar Catalogue (https://www.atnf.csiro.au/people/pulsar/psrcat/; Manchester et al. 2005). This work made use of the Starlink software (Currie et al. 2014), which is currently supported by the East Asian Observatory. We thank the anonymous reviewer for the useful comments that improved this paper. T.-X.L. and P.Z. acknowledge the support from NSFC grant No. 12273010. H.-N.He is supported by Project for Young Scientists in Basic Research of Chinese Academy of Sciences (No. YSBR-061), and by NSFC under the grants No. 12173091 and No. 12333006.

*Software:* astropy (Astropy Collaboration et al. 2013, 2018), Starlink (Currie et al. 2014).

## Appendix A
## Pulsar Catalogue

Table 2 lists information about 44 known pulsars located within $26\overset{\circ}{.}6 < l < 30\overset{\circ}{.}6$, $|b| \leqslant 1\overset{\circ}{.}25$ from the ATNF Pulsar Catalogue.

Table 2
Pulsar Catalogue

| Name | GLong (°) | GLat (°) | Distance (kpc) | Age (Yr) | Ref. | Assoc.[a] |
|---|---|---|---|---|---|---|
| J1841−0524 | 27.024 | −0.333 | 4.12 | $3.02 \times 10^4$ | Hobbs et al. (2004) | |
| J1838−0453 | 27.070 | 0.708 | 6.63 | $5.19 \times 10^4$ | Morris et al. (2002) | |
| B1841−05 | 27.073 | −0.941 | 5.40 | $4.17 \times 10^5$ | Clifton & Lyne (1986) | |
| J1839−0459 | 27.147 | 0.321 | 3.94 | $2.80 \times 10^6$ | Morris et al. (2002) | |
| B1834−04 | 27.167 | 1.130 | 4.36 | $3.38 \times 10^6$ | Clifton & Lyne (1986) | |
| J1841−0500 | 27.323 | −0.034 | 4.97 | $4.16 \times 10^5$ | Camilo et al. (2012) | |
| J1841−0456 | 27.387 | −0.007 | 9.60 | $4.57 \times 10^3$ | Vasisht & Gotthelf (1997) | Kes73 |
| J1843−0510 | 27.391 | −0.520 | 4.07 | $2.73 \times 10^6$ | Ng et al. (2015) | |
| J1839−0436 | 27.407 | 0.654 | 4.48 | $2.92 \times 10^6$ | Lorimer et al. (2006) | G27.8+0.6? |
| J1840−0445 | 27.486 | 0.203 | 4.56 | $5.93 \times 10^5$ | Bates et al. (2012) | |
| J1841−04 | 27.490 | 0.090 | 0.99 | … | Tyul'bashev et al. (2018) | |
| J1843−0459 | 27.581 | −0.505 | 4.89 | $1.40 \times 10^7$ | Morris et al. (2002) | |
| J1844−0502 | 27.670 | −0.769 | 4.58 | $8.60 \times 10^7$ | Hobbs et al. (2004) | |
| J1844−0452 | 27.752 | −0.576 | 5.68 | $6.28 \times 10^6$ | Hobbs et al. (2004) | |
| B1838−04 | 27.818 | 0.279 | 4.40 | $4.61 \times 10^5$ | Clifton & Lyne (1986) | G27.8+0.6? |
| J1839−0402 | 28.016 | 0.730 | 4.23 | $1.07 \times 10^6$ | Morris et al. (2002) | G27.8+0.6? |
| J1842−0415 | 28.086 | 0.111 | 3.60 | $3.80 \times 10^5$ | Lorimer et al. (2000) | |
| B1841−04 | 28.096 | −0.548 | 3.07 | $4.01 \times 10^6$ | Clifton et al. (1987) | |
| B1842−04 | 28.193 | −0.785 | 4.09 | $6.81 \times 10^5$ | Clifton et al. (1987) | |
| B1839−04 | 28.347 | 0.174 | 3.71 | $5.74 \times 10^7$ | Clifton & Lyne (1986) | G28.36+0.21? |
| J1843−0408 | 28.374 | −0.172 | 3.98 | $5.18 \times 10^6$ | Hobbs et al. (2004) | |
| J1841−0345 | 28.424 | 0.437 | 3.78 | $5.59 \times 10^4$ | Lorimer et al. (2000) | |
| J1839−0332 | 28.462 | 0.934 | 4.04 | $8.91 \times 10^6$ | Ng et al. (2015) | |
| J1843−0355 | 28.484 | 0.056 | 5.80 | $2.02 \times 10^6$ | Morris et al. (2002) | |
| J1847−0427 | 28.487 | −1.120 | 3.97 | $6.96 \times 10^8$ | Ng et al. (2015) | |
| J1839−0321 | 28.601 | 1.095 | 7.85 | $3.02 \times 10^5$ | Morris et al. (2002) | |
| J1844−0346 | 28.787 | −0.191 | … | $1.16 \times 10^4$ | Clark et al. (2017) | |
| B1844−04 | 28.876 | −0.939 | 3.42 | $1.83 \times 10^5$ | Davies et al. (1970) | |
| J1841−0310 | 28.968 | 0.776 | 4.17 | $7.83 \times 10^7$ | Hobbs et al. (2004) | |
| J1842−0309 | 29.078 | 0.584 | 8.64 | $1.42 \times 10^6$ | Hobbs et al. (2004) | |
| J1844−0310 | 29.343 | 0.036 | 5.97 | $8.13 \times 10^5$ | Lorimer et al. (2000) | G29.38+0.10? |
| J1845−0316 | 29.390 | −0.255 | 5.06 | $3.71 \times 10^5$ | Lorimer et al. (2000) | G29.41−0.18? |
| J1844−0302 | 29.395 | 0.243 | 5.26 | $2.43 \times 10^6$ | Ng et al. (2015) | |
| J1844−03 | 29.523 | 0.074 | … | … | Torii et al. (1998) | G29.6+0.1? |
| J1844−0256 | 29.574 | 0.119 | 6.03 | … | Crawford et al. (2002) | G29.6+0.1? |
| J1846−0257 | 29.707 | −0.198 | 4.00 | $4.42 \times 10^5$ | McLaughlin et al. (2006) | |
| J1846−0258 | 29.712 | −0.240 | 5.80 | 728 | Gotthelf et al. (2000) | Kes75 |
| B1842−02 | 29.727 | 0.235 | 4.89 | $4.81 \times 10^5$ | Clifton & Lyne (1986) | |
| J1849−0317 | 29.834 | −1.173 | 1.21 | $4.81 \times 10^5$ | Morris et al. (2002) | |
| J1843−0211 | 30.084 | 0.768 | 5.93 | $2.22 \times 10^6$ | Hobbs et al. (2004) | |
| J1841−0157 | 30.099 | 1.216 | 7.93 | $5.81 \times 10^5$ | Hobbs et al. (2004) | |
| J1842−0153 | 30.283 | 1.022 | 6.88 | $2.48 \times 10^6$ | Morris et al. (2002) | |
| J1851−0241 | 30.515 | −1.186 | 7.92 | $8.66 \times 10^5$ | Hobbs et al. (2004) | |
| J1843−0137 | 30.543 | 1.087 | 7.66 | $4.30 \times 10^6$ | Hobbs et al. (2004) | |

**Note.**
[a] The pulsars that are spatially located in the circle region of SNRs or SNR candidates are listed here. The "?" denotes it is uncertain if this SNR or SNR candidate is associated with the corresponding pulsar.

## Appendix B
## Nonthermal Point Sources

The coordinate information and type of the nonthermal point sources detected in Figure 2 are shown in Table 3. Almost all of the nonthermal point sources detected in our spectral index map are radio sources that have not been identified yet. Two point sources have three counterparts. [JD2012] G027.375+01.170 is an extragalactic H II (Jones & Dickey 2012) so it may not correspond to nonthermal point source No. 2. The three counterparts of source No. 4 are not exactly in the position of the point source in the high-resolution THOR+VGPS data. Hence, the correspondence between them is relatively weak.

Table 3
Nonthermal Point Sources

| Number | Name | GLong (°) | GLat (°) | Type |
|---|---|---|---|---|
| 1 | 4C-06.55 | 26.687 | −1.028 | Radio Source |
| 2 | [JD2012] G027.375+01.170 | 27.375 | 1.170 | H II |
|   | RFS 440 | 27.378 | 1.163 | Radio Source |
|   | PMN J1837-0424 | 27.378 | 1.169 | Radio Source |
| 3 | NVSS J183847-040042 | 27.920 | 0.977 | Radio source |
| 4 | RFS 453 | 28.130 | −1.017 | Radio Source |
|   | IRAS 18435-0446 | 28.136 | −0.990 | Star |
|   | PMN J1846-0445 | 28.136 | −1.020 | Radio Source |
| 5 | NVSS J183909-031823 | 28.589 | 1.218 | Radio Source |
| 6 | PMN J1839-0300 | 28.905 | 1.247 | Radio Source |
| 7 | GPSR 029.089+0.511 | 29.090 | 0.511 | Radio Source |
| 8 | GPSR 029.228+0.518 | 29.228 | 0.517 | Radio Source |
| 9 | NVSS J184230-021839 | 29.856 | 0.933 | Radio Source |

## ORCID iDs

Tian-Xian Luo https://orcid.org/0000-0002-9911-2509
Ping Zhou https://orcid.org/0000-0002-5683-822X
Hao-Ning He https://orcid.org/0000-0002-8941-9603